\documentstyle[12pt,epsfig]{article}

\newcommand{\be}{\begin{eqnarray}}
\newcommand{\ee}{\end{eqnarray}}
\newcommand{\e}{\epsilon}
\renewcommand{\l}{\lambda}
\renewcommand{\Re}{{\rm Re}}
\renewcommand{\Im}{{\rm Im}}

\begin{document}

\newpage
\setcounter{page}{0}

\begin{titlepage}
 \begin{flushright}
 \hfill{\bf (hep-ph/9911281)}\\
 \hfill{YUMS 99--026}\\

 \end{flushright}
\vspace*{1.0cm}

\begin{center}
{\large\bf Search for New Physics  \\ 
in the Semileptonic $D_{l4}$ Decays, $D^\pm \to K\pi l^\pm \nu$ }
\end{center}
\vskip 0.8cm
\begin{center}
{\sc C. S.~Kim$^{\mathrm{a,}}$\footnote{e-mail: kim@cskim.yonsei.ac.kr,
~~ http://phya.yonsei.ac.kr/\~{}cskim/},
 Jake Lee$^{\mathrm{a,}}$\footnote{e-mail: jilee@theory.yonsei.ac.kr} and 
 W.~Namgung$^{\mathrm{b,}}$\footnote{e-mail: ngw@cakra.dongguk.ac.kr}}

\vskip 0.5cm

\begin{small} 
$^{\mathrm{a}}$ {\it Department of Physics, Yonsei University, 120-749 Seoul, 
                  Korea} \\
\vskip 0.2cm
$^{\mathrm{b}}$ {\it Department of Physics, Dongguk University, 100-715 
                  Seoul, Korea}
\end{small}
\end{center}

\vspace{0.5cm}
\begin{center}
 (\today)
\end{center}

\setcounter{footnote}{0}
\vspace*{1.0cm}

\begin{abstract}

\noindent
New physics effects through the direct CP violation and the decay rate change
are investigated
in the semileptonic $D_{l4}$ decays, $D^\pm\to K\pi l^\pm \nu$,
by including a scalar-exchange interaction with a  complex coupling. 
In the decay process, we included various excited states as 
intermediate states decaying to the final hadrons, $K+\pi$,
and found that among the intermediate states only the lowest
state ($K^*$) is dominant and the other higher excited states are
negligible, contrary to the $B_{l4}$ decays. 
We also obtained constraints on the new complex coupling
within the multi-Higgs doublet model and the scalar leptoquark models.
\vskip 1cm
\noindent
PACS numbers: 13.20.Fc, 12.60.-i, 11.30.Er
\end{abstract}
\vskip 1cm
\end{titlepage}

\newpage
\baselineskip .29in
\renewcommand{\thefootnote}{\alph{footnote}}

\section{Introduction}

Recently we studied the possibility of probing the direct CP violation in
the $B_{l4}$ decays, $B^\pm\to D\pi l^\pm \nu$ \cite{bl4} and
$B^\pm\to \pi\pi l^\pm \nu$ \cite{bpi}, 
in extensions of the Standard Model (SM)
by including  a scalar-exchange interaction
with a complex coupling in the weak charged current. 
There we  considered, as specific models, the
multi-Higgs doublet (MHD) model and the scalar-leptoquark (SLQ) models.
And we investigated mainly  the direct CP violation effects, {
\it i.e.},  CP-odd asymmetries for maximally-allowed
values of the imaginary part of the additional scalar coupling,
even though we also found that such a scalar coupling could induce a sizable 
effect to the CP-even total decay rate.
In the present paper, we take a more general approach 
and investigate new physics
effects by considering changes in the (CP-even) total decay rate as well as
the (CP-odd) CP violation effects in $D_{l4}$ decays, 
$D^\pm\to K\pi l^\pm \nu$.
Here we  find constraints to both real and imaginary parts of the new
scalar coupling, 
by comparing our predictions of the decay rate and the direct CP 
violation effects with the observable experimental results.  

As is well known, in order to observe the direct CP violation effects, 
there should
exist interferences not only through weak CP-violating phases but also with 
different CP-conserving strong phases.
In the decay  $D^\pm\to K\pi l^\pm \nu$, we consider it as a two-step process: 
$D \to (\sum_{i} \tilde{K}_i) l\nu \to K\pi l\nu $, 
where $\tilde{K}_i$ stands for an intermediate
state which  decays to $K+ \pi$.
In this picture CP-conserving phases come from the absorptive parts
of the intermediate resonances.
The relevant resonance states are $K^*$, $K_0^*$ and $K_2^*$ mesons,
which decay dominantly to $K\pi$ mode (see Table.~1).
As shown in Refs. \cite{bl4,bpi} for the $B_{l4}$ decays, 
the inclusion of higher excited
states could amplify the CP violation effects. However, we now find that
this is not true anymore for the $D_{l4}$ decays, because
in the $D_{l4}$ decays final state phase spaces are much smaller 
for those higher excited states than
the available phase spaces in the $B_{l4}$ decays. 
Therefore, we anticipate that the effects of higher 
excited resonances will be correspondingly reduced 
and  the dominant contributions will
come mainly through the lowest state $K^*(892)$.
We will discuss more on this later.

In Section 2, we present in detail our formalism for the 
$D^\pm\to K\pi l^\pm \nu$ decays within the SM and in the models beyond the SM.
Section 3 is devoted to the numerical analyses, and concluding remarks are 
also in Section 3.
\begin{table}
{Table~1}. {Properties and branching ratios of $K\pi$ resonances}\par
\begin{center}
\begin{tabular}{|c|l|c|c|c|c|}
\hline
Label $i$ & $\qquad\quad \tilde{K}_i$ & $J^P$ & $m_i$ (MeV) & $\Gamma_i$ (MeV) &
${\cal BR}_i(\tilde{K}_i\to K^+ \pi^-)$\\
\hline
$0$ & $\tilde{K}_0=K_0^*(1430)$ & $0^+$ & $1429$ & $287$ & 0.62\\
$1$ & $\tilde{K}_1=K^*(892)$ & $1^-$ & $896$ & $51$ & 0.67\\
$2$ & $\tilde{K}_2=K_2^* (1430)$ & $2^+$ & $1432$ & $109$ & 0.33\\
\hline
\end{tabular}
\end{center}
\end{table}

\section{Theoretical Details of the $D_{l4}$ Decays }

We first describe the formalism within the SM for the decay 
$D^- \to K\pi l^- \nu$.
Its extensions to the models beyond the SM are obtained by
rather simple appropriate modifications.
The decay amplitudes for the processes of Fig.~1
\be
D^-(p_D)\to \tilde{K}_i(p_i,\l_i)+W^*(q)\to
 K^+(p_K)+\pi^-(p_\pi)+l^-(p_l,\l_l)+\bar{\nu}(p_\nu)
\ee
are expressed as
\be
{\cal A}^{\lambda_l}&=&-V_{cs}\frac{G_F}{\sqrt{2}}\sum_{i}\sum_{\lambda_i}
         \langle l^-(p_l,\lambda_l)\bar{\nu}(p_\nu)|j^{\mu\dagger}|0\rangle 
             \langle \tilde{K}_i(p_i,\lambda_i)|J_{\mu}|D^-(p_D)\rangle \nonumber\\
         & &\times \Pi_i(s_{_M})\langle K^+(p_K)\pi^-(p_{\pi})\|\tilde{K}_i(p_i,\lambda_i)\rangle ,
\ee
where $\l_i=0$ for spin $0$ states ($K_0^*$),
$\l_i=\pm 1,0$ for spin $1$ states ($K^*$),
$\l_i=\pm 2,\pm 1,0$ for spin $2$ states ($K_2^*$),
and $\l_l$ is the lepton helicity, $\pm\frac{1}{2}$.

\begin{figure}[ht]
\hbox to\textwidth{\hss\epsfig{file=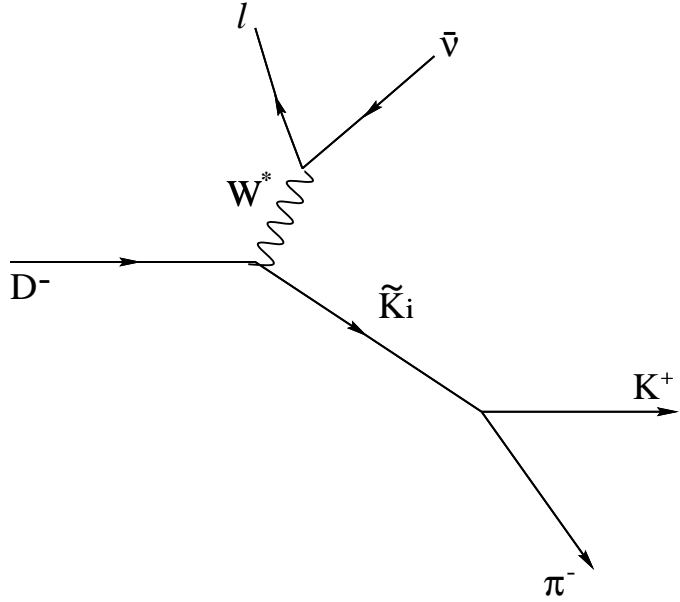,height=9cm,width=8cm,angle=-90}\hss}
\caption{Diagrams for $D^-\to \tilde{K}_iW^*\to K^+\pi^- l^-\bar{\nu}_l$ decays
within the SM.}
\label{fig:diagram}
\end{figure}

The leptonic and hadronic currents are defined, respectively, as
\be
j^\mu &=& \bar{\psi}_\nu \gamma^\mu(1-\gamma_5)\psi_l,\nonumber\\
J^\mu &=& \bar{\psi}_c \gamma^\mu(1-\gamma_5)\psi_s.
\ee
We assume that the resonance contributions of the intermediate states
can be treated by the Breit-Wigner form, which is written
in the narrow width approximation as
\be
\Pi_i (s_{_M})=\frac{\sqrt{m_i\Gamma_i/\pi}}{s_{_M}-m_i^2+im_i\Gamma_i},
\ee
where $s_{_M}=(p_K + p_{\pi})^2$ and the $m_i$'s and $\Gamma_i$'s are the masses
and widths of the resonances, respectively (see Table 1).
For the decay parts of the resonances we use \cite{cancel}
\be
\langle K^+(p_K)\pi^-(p_{\pi})\|\tilde{K}_i(p_i,\lambda_i)\rangle 
=\sqrt{{\cal BR}_i}Y^{\lambda_i}_{\lambda_i max} (\theta^*,\phi^*),
\ee
where $Y^m_l(\theta,\phi)$ are the $J=l$ spherical harmonics, 
and the angles $\theta^*$
and $\phi^*$ are those of the final state $\pi^-$ specified in the $\tilde{K}_i$
rest frame, as defined in Fig. 2c.
The couplings of $\tilde{K}_i$ to $K\pi$ are effectively taken into account by
the branching fractions, ${\cal BR}_i(\tilde{K}_i\to K^+\pi^-)$.

\begin{figure}[tb]
\hbox to\textwidth{\hss\epsfig{file=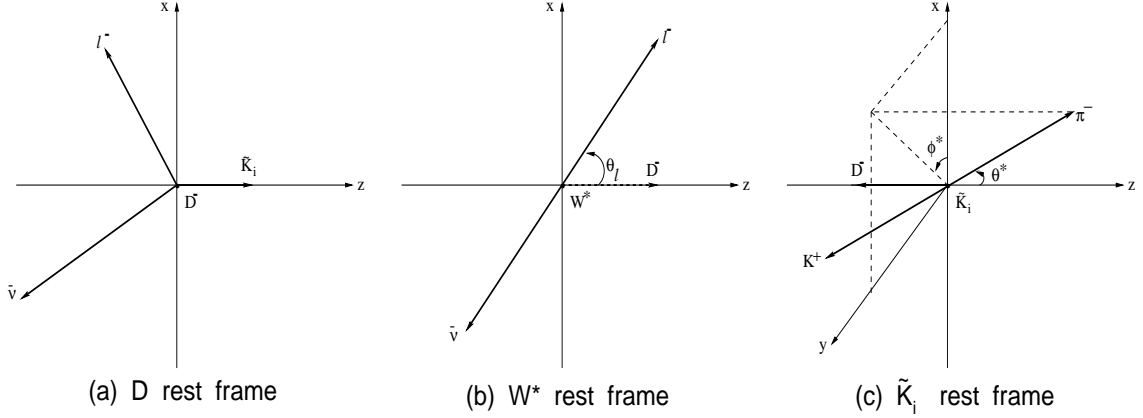,height=15cm,width=5.5cm,angle=-90}\hss}
\caption{The decay $D^-\to \tilde{K}_iW^*\to (K^+\pi^-)(l\bar{\nu})$ viewed from the
(a) $D^-$, (b) $W^*$ and (c) $\tilde{K}_i$ rest frames.}
\label{fig:frame}
\end{figure}

In order to obtain the full helicity amplitude of the 
$D^- \to K\pi l^- \nu$ decay,
we first consider the amplitude of $D^-\to \tilde{K}_i l^- \bar{\nu}_l$,
denoted as ${\cal M}^{\l_l}_{\l_i}$:
\be
{\cal M}^{\l_l}_{\l_i}=-V_{cs}\frac{G_F}{\sqrt{2}}
         \langle l^-(p_l,\lambda_l)\bar{\nu}(p_\nu)|j^{\mu\dagger}|0\rangle 
             \langle \tilde{K}_i(p_i,\lambda_i)|J_{\mu}|D^-(p_D)\rangle .
\ee
We express the matrix elements
${\cal M}^{\l_l}_{\l_i}$ into the following form:
\be
{\cal M}^{\l_l}_{\l_i}=V_{cs}\frac{G_F}{\sqrt{2}}
       \sum_{\l_{_W}}\eta_{\l_{_W}}L^{\l_l}_{\l_{_W}}H^{\l_i}_{\l_{_W}},
\label{smamp}
\ee
where for the decays $D \to \tilde{K}_i W^*$ and $W^* \to l \bar{\nu}$, respectively,
\be
H^{\l_i}_{\l_{_W}}&=&\e^*_{W\mu}\langle \tilde{K}_i(p_i,\l_i)|J^\mu|D^-(p_D)\rangle ,\nonumber\\
L^{\l_l}_{\l_{_W}}&=&\e_{W\mu}\langle l^-(p_l,\l_l)\bar{\nu}(p_\nu)|j^{\mu\dagger}|0\rangle ,
\ee
in terms of the polarization vectors $\e_{_W} \equiv \e(q,\l_{_W})$ of 
the virtual $W$.
These $\e_{_W}$'s satisfy the relation 
\be
-g^{\mu\nu}=\sum_{\l_{_W}}\eta_{\l_{_W}}\e^\mu_{_W} \e^{*\nu}_{_W},
\label{metric}
\ee
where the summation is over the helicities  $\l_{_W} =\pm 1,0,s$ of the virtual $W$, 
with the metric $\eta_\pm=\eta_0=-\eta_s=1$.

We evaluate the leptonic amplitude $L^{\l_l}_{\l_{_W}}$ in the rest frame of
the virtual $W^*$ (see Fig. 2b).
Using the 2-component spinor formalism \cite{hagiwara} with
explicit polarization vectors \cite{bpi}, we find
\be
&&L^-_\pm=2\sqrt{q^2}vd_\pm,\; L^-_0=-2\sqrt{q^2}vd_0,\; \hspace{1.1cm} L^-_s=0,\nonumber\\
&&L^+_\pm=\pm 2m_lvd_0,\; L^+_0=\sqrt{2}m_lv(d_+-d_-),\; L^+_s=-2m_lv,
\label{Lpm}
\ee
where
\be
v=\sqrt{1-\frac{m_l^2}{q^2}},\; d_\pm=\frac{1\pm\cos\theta_l}{\sqrt{2}},\;
~~{\rm and}~~d_0=\sin\theta_l.
\ee
Here we show only the sign of $\l_l$ as a superscript on $L$.
Note that the $L^+$ amplitudes are proportional
to the lepton mass $m_l$, and the scalar amplitude $L^-_s$ vanishes due to
angular momentum conservation.

For the $D\to \tilde{K}_i$ transition through the weak charged current 
\be
J^\mu=V^\mu-A^\mu,
\ee
the most general forms of matrix elements are,
\be
{\rm for}&K_0^*(0^+)&{\rm states}:\nonumber\\
&&\langle K_0^*(p_i)|V_\mu|D(p_D)\rangle =0,\nonumber\\
&&\langle K_0^*(p_i)|A_\mu|D(p_D)\rangle 
 =u_+(q^2)(p_D+p_i)_\mu+u_-(q^2)(p_D-p_i)_\mu;\nonumber\\
{\rm for}&K^*(1^-)&{\rm states}:\nonumber\\
&&\langle K^*(p_i,\e_1)|V_\mu|D(p_D)\rangle 
 =ig(q^2)\e_{\mu\nu\rho\sigma}\e_1^{*\nu}(p_D+p_i)^\rho
                           (p_D-p_i)^\sigma,\nonumber\\
&&\langle K^*(p_i,\e_1)|A_\mu|D(p_D)\rangle 
 =f(q^2)\e^*_{1\mu}+a_+(q^2)(\e_1^*\cdot p_D)(p_D+p_i)_\mu\nonumber\\
    &&\hspace{4.5cm} +a_-(q^2)(\e_1^*\cdot p_D)(p_D-p_i)_\mu;\nonumber\\
{\rm for}&K_2^*(2^+)&{\rm states}:\nonumber\\
&&\langle K_2^*(p_i,\e_2)|V_\mu|D(p_D)\rangle 
 =ih(q^2)\e_{\mu\nu\l\rho}\e_2^{*\nu\alpha}p_{D\alpha}
                  (p_D+p_i)^\l(p_D-p_i)^\rho,\nonumber\\
&&\langle D_2^*(p_i,\e_2)|A_\mu|D(p_D)\rangle 
 =k(q^2)\e^*_{2\mu\nu}p_D^\nu +b_+(q^2)(\e^*_{2\alpha\beta}
                   p_D^\alpha p_D^\beta)(p_D+p_i)_\mu\nonumber\\
 &&\hspace{4.5cm} +b_-(q^2)(\e^*_{2\alpha\beta}p_D^\alpha p_D^\beta)(p_D-p_i)_\mu,
\label{BMamp}
\ee
where $\e_1$ and $\e_2$ are the polarization vectors of the spin 1 
and spin 2 states, respectively.
Using the above expressions and the polarization vectors \cite{bpi}, 
we find the non-zero $D\to \tilde{K}_iW^*$ amplitudes,
\be
{\rm for}&i=0,&H^0_{\l_{_W}}\equiv S^0_{\l_{_W}},\nonumber\\
&&S^0_0=-u_+(q^2)\frac{\sqrt{Q_+Q_-}}{\sqrt{q^2}},\nonumber\\
&&S^0_s=-\left(u_+(q^2)\frac{(m_D^2-s_{_M})}{\sqrt{q^2}}+u_-(q^2)\sqrt{q^2}\right),\\
{\rm for}&i=1^{^{(\prime)}},&H^{\l_1}_{\l_{_W}}\equiv V^{\l_1}_{\l_{_W}},\nonumber\\
&&V^0_0=-\frac{1}{2\sqrt{s_{_M}q^2}}
 \left[f(q^2)(m_D^2-s_{_M}-q^2)+a_+(q^2)Q_+Q_-\right],\nonumber\\
&&V^{\pm 1}_{\pm 1}=f(q^2)\mp g(q^2)\sqrt{Q_+Q_-},\nonumber\\
&&V^0_s=-\frac{\sqrt{Q_+Q_-}}
 {2\sqrt{s_{_M}q^2}}\left[f(q^2)+a_+(q^2)(m_D^2-s_{_M})+a_-(q^2)q^2\right],\\
{\rm for}&i=2,&H^{\l_2}_{\l_{_W}}\equiv T^{\l_2}_{\l_{_W}},\nonumber\\
&&T^0_0=-\frac{1}{2\sqrt{6}}\frac{\sqrt{Q_+Q_-}}{s_{_M}\sqrt{q^2}}
         \left[k(q^2)(m_D^2-s_{_M}-q^2)+b_+(q^2)Q_+Q_-\right],\nonumber\\
&&T^{\pm 1}_{\pm 1}=\frac{1}{2\sqrt{2}}\sqrt{\frac{Q_+Q_-}{s_{_M}}}
          [k(q^2)\mp h(q^2)\sqrt{Q_+Q_-}],\nonumber\\
&&T^0_s=-\frac{1}{2\sqrt{6}}\frac{Q_+Q_-}{s_{_M}\sqrt{q^2}}
         \left[k(q^2)+b_+(q^2)(m_D^2-s_{_M})+b_-(q^2)q^2\right],
\ee
where 
\be
Q_\pm=(m_D\pm\sqrt{s_{_M}})^2-q^2.
\label{Qpm}
\ee
Combining all the formulae, we can write the SM helicity amplitudes of
$D^-\to K^+\pi^- l^-\bar{\nu}$ decays as
\be
{\cal A}^{\lambda_l}&=&V_{cs}\frac{G_F}{\sqrt{2}}\bigg[
      \sum_{\l=0,s} \eta_\l L^{\l_l}_\l (\Pi_{K_0^*} S^0_\l Y^0_0 +
      +\Pi_{K^*} V^0_\l Y^0_1
        +\Pi_{K_2^*} T^0_\l Y^0_2)\nonumber\\
      &&+\sum_{\l=\pm 1} L^{\l_l}_\l (\Pi_{K^*} V^\l_\l Y^\l_1
        +\Pi_{K_2^*} T^\l_\l Y^\l_2)\bigg],
\label{amp}
\ee
where
\be
\Pi_{K_0^*}&\equiv&\sqrt{{\cal BR}_0}\Pi_0\nonumber\\
\Pi_{K^*}&\equiv&\sqrt{{\cal BR}_1}\Pi_1\nonumber\\
\Pi_{K_2^*}&\equiv&\sqrt{{\cal BR}_2}\Pi_2.
\ee
The differential partial width  can be expressed as
\be
d\Gamma(D^- \to K^+ \pi^- l^- \bar{\nu}_l)
 =\frac{1}{2m_D}\sum_{\l_l}|{\cal A}^{\l_l}|^2
       \frac{(q^2-m_l^2)\sqrt{Q_+Q_-}}{256\pi^3m_D^2q^2}d\Phi_4,       
\ee
where the 4 body phase space $d \Phi_4$ is
\be
d \Phi_4 \equiv  ds_{_M} \cdot dq^2 \cdot d\cos\theta^* \cdot 
                 d\cos\theta_l \cdot d\phi^*.
\ee

So far we have established the SM formalism for the $D_{l4}$ decays.
Now we extend
the virtual $W$-exchange part in Fig.~1 by including an additional scalar
interaction with the complex coupling.
Then, the decay amplitudes for $D^- \to K^+ \pi^- l^- \bar{\nu}_l$ can be 
expressed as
\be
{\cal A}^{\lambda_l}&=&-V_{cs}\frac{G_F}{\sqrt{2}}\sum_{i}\sum_{\lambda_i}
     \bigg[\langle l^-(p_l,\lambda_l)\bar{\nu}(p_\nu)|j^{\mu\dagger}|0\rangle 
             \langle \tilde{K}_i(p_i,\lambda_i)|J_{\mu}|D^-(p_D)\rangle \nonumber\\
        & &+\zeta\langle l^-(p_l,\lambda_l)\bar{\nu}(p_\nu)|j_s^{\dagger}|0\rangle 
             \langle \tilde{K}_i(p_i,\lambda_i)|J_s|D^-(p_D)\rangle \bigg]\nonumber\\
         & &\times \Pi_i(s_{_M})\langle K^+(p_K)\pi^-(p_{\pi})\|\tilde{K}_i(p_i,\lambda_i)\rangle ,
\ee
where the scalar currents are
\be
j_s=\bar{\psi}_\nu(1-\gamma_5)\psi_l,\;\;
J_s=\bar{\psi}_c(1-\gamma_5)\psi_s.
\ee
Here the parameter $\zeta$, which parameterizes contributions from physics beyond
the SM, is in general a complex number.
By using the Dirac equation for the leptonic current, $q_\mu j^\mu=m_l j_s$,
the amplitude can be written as
\be
{\cal A}^{\lambda_l}&=&-V_{cs}\frac{G_F}{\sqrt{2}}\sum_{i}\sum_{\lambda_i}
     \langle l^-(p_l,\lambda_l)\bar{\nu}(p_\nu)|j^{\mu\dagger}|0\rangle 
             \langle \tilde{K}_i(p_i,\lambda_i)|\Omega_\mu|D^-(p_D)\rangle \nonumber\\
         & &\times \Pi_i(s_{_M})\langle K^+(p_K)\pi^-(p_{\pi})\|\tilde{K}_i(p_i,\lambda_i)\rangle ,
\label{dmamp}
\ee
where the effective hadronic current $\Omega_\mu$ is defined as
\be
\Omega_\mu\equiv J_\mu + \zeta \frac{q_\mu}{m_l} J_s.
\ee
In this case the amplitudes ${\cal M}^{\l_l}_{\l_i}$ of $D\to \tilde{K}_i l\bar{\nu}$
have the same form as the previous SM case (\ref{smamp}) except for the 
modification in the hadronic current part due to the additional scalar current:
\be
{\cal M}^{\l_l}_{\l_i}=\frac{G_F}{\sqrt{2}}V_{cs}
       \sum_{\l_{_W}}\eta_{\l_{_W}}L^{\l_l}_{\l_{_W}}{\cal H}^{\l_i}_{\l_{_W}},
\ee
where ${\cal H}^{\l_i}_{\l_{_W}}$ stands for the hadronic amplitudes modified 
by the scalar current $J_s$.
Using the equations of motion for $c$ and $s$ quarks, we get within the on-shell
approximation
\be
J_s = (p_c^\mu-p_s^\mu) \left[\frac{V_\mu}{m_c-m_s}-\frac{A_\mu}{m_c+m_s}\right] .
\ee
Later we use the approximation, 
$(p_c^\mu-p_s^\mu) \approx (p_D^\mu-p_{\tilde{K}_i}^\mu) \equiv q^\mu$,
which has been generally assumed in quark model calculations of the form factors.
After explicit calculation, we find that the additional scalar current modifies
only the scalar component of ${\cal H}^{\l_i}_{\l_{_W}}$: 
\be
{\cal H}^0_s&=&(1+\zeta^\prime) H^0_s,\nonumber\\
{\rm and}~~~{\cal H}^{\l_i}_{\l_{_W}}&=&H^{\l_i}_{\l_{_W}}
        \;\;{\rm for}\;\; \l_{_W} = 0,~\pm 1,
\ee
where
\be
\zeta^\prime=\frac{q^2}{m_l(m_c+m_s)}\zeta .
\label{zetap}
\ee

So far, we constructed a formalism for $D^-_{l4}$ decays.
Since the initial $D^-$ system is not CP self-conjugate, any genuine
CP-odd observable can be constructed by considering both the $D^-$
decay and its charge-conjugated $D^+$ decay, and by identifying the CP
relations of their kinematic distributions.
Before constructing possible CP-odd asymmetries explicitly, we calculate
the decay amplitudes for the charge-conjugated process
$D^+\to K^- \pi^+ l^+ \nu_l$. 
For the charge-conjugated $D^+$ decays,
the amplitudes can be written as
\be
{\bar{\cal A}}^{\l_l}&=&-V_{cs}^*\frac{G_F}{\sqrt{2}}\sum_{i}\sum_{\l_i}
   \langle l^+(p_l,\lambda_l)\nu(p_\nu)|j^{\mu}|0\rangle 
    \langle \overline{\tilde{K}}_i(p_i,\lambda_i)|\Omega^\dagger_{\mu}|D^+(p_D)\rangle \nonumber\\
  & &\times \Pi_i(s_{_M})\langle K^-(p_K)\pi^+(p_{\pi})\|\overline{\tilde{K}}_i(p_i,\lambda_i)\rangle .
\label{dpamp}
\ee
Similarly to the $D^-$ decay,
the leptonic amplitudes $\bar{L}^{\l_l}_{\l_{_W}}$ for $D^+$ decay are
\be
&&\bar{L}^+_\pm=-2\sqrt{q^2}vd_\mp,\; \bar{L}^+_0
   =-2\sqrt{q^2}vd_0,\; \hspace{1.1cm} \bar{L}^+_s=0,\nonumber\\
&&\bar{L}^-_\pm=\pm 2m_lvd_0,\;\;\;\; \bar{L}^-_0=\sqrt{2}m_lv(d_+-d_-),\; \bar{L}^-_s=-2m_lv.
\ee
And the transition amplitudes $\overline{{\cal H}}^{\l_i}_{\l_{_W}}$ 
for $D^+\to \overline{\tilde{K}}_iW^*$
are given by a simple modification of the 
amplitudes ${\cal H}^{\l_i}_{\l_{_W}}$ of the $D^-$ decays:
\be
\overline{{\cal H}}^{\l_i}_{\l_{_W}}={\cal H}^{\l_i}_{\l_{_W}}
    \{g\to -g,\;h\to -h,\;f_\pm\to-f_\pm\; ; \zeta\to \zeta^*\},
\ee
which is  expected from the property that vector currents change sign
under the charge conjugation.

It is easy to see that
if $\zeta$ is real, the amplitude (\ref{dmamp}) of the $D^-$ decay and
(\ref{dpamp}) of the $D^+$ decay satisfy the CP transition relation:
\be
{\cal A}^{\pm}(\theta^*,\phi^*,\theta_l)
   =\eta_{CP}{\bar{\cal A}}^{\mp}(\theta^*,-\phi^*,\theta_l),
\label{cprel}
\ee
where $\theta^*$ and $\phi^*$ in $\bar{\cal A}^{\l_l}$ are the angles of
the final state $\pi^+$, while those in ${\cal A}^{\l_l}$ are for $\pi^-$.  
Then, with a complex phase $\zeta$, 
$d\Gamma/d\Phi_4$ can be decomposed into a CP-even part ${\cal S}$ and
a CP-odd part ${\cal D}$:
\be
\frac{d\Gamma}{d\Phi_4}=\frac{1}{2}({\cal S}+{\cal D}).
\ee
The CP-even part ${\cal S}$ and the CP-odd part ${\cal D}$ can be easily
identified by making use of the CP relation (\ref{cprel}) between $D^-$ and 
$D^+$ decay amplitudes, and they are expressed as
\be
{\cal S} =\frac{d(\Gamma+\overline{\Gamma})}{d\Phi_4},\quad
{\cal D} =\frac{d(\Gamma-\overline{\Gamma})}{d\Phi_4},
\ee
where $\Gamma$ and $\overline{\Gamma}$ are the decay rates for $D^-$ and $D^+$, 
respectively. And 
we use the same kinematic variables $\{s_{_M},q^2,\theta^*,\theta_l\}$
for the $d\overline{\Gamma}/d\Phi_4$ except for the replacement of 
${\phi^*} \to -\phi^*$, 
as shown in Eq. (\ref{cprel}).
The CP-even ${\cal S}$ term and the CP-odd ${\cal D}$ term can be obtained from
$D^\mp$ decay probabilities.
The CP-even quantity ${\cal S}$ is
\be
{\cal S}=2C(q^2,s_{_M})\Sigma,
\ee
with
\be
&&\Sigma=(L^-_0S^0_0Y^0_0)^2|\Pi_{K_0^*}|^2+|\langle V^-\rangle \Pi_{K^*}|^2+|\langle T^-\rangle \Pi_{K_2^*}|^2\nonumber\\
   &+&2(L^-_0S^0_0Y^0_0)\Re(\Pi_{K_0^*}\Pi_{K^*}^*\langle V^-\rangle^*+\Pi_{K_0^*}\Pi_{K_2^*}^*\langle T^-\rangle^*)+2\Re(\Pi_{K^*}\Pi_{K_2^*}^*\langle V^-\rangle \langle T^-\rangle^*)\nonumber\\
   &+&|\Pi_{K_0^*}|^2|L^+_0S^0_0Y^0_0-(1+\zeta^\prime)L^+_sS^0_sY^0_0|^2\nonumber\\
   &+&|\Pi_{K^*}|^2[|\langle V^+\rangle|^2 +(L^+_sV^0_sY^0_1)^2|1+\zeta^\prime|^2
      -2(L^+_sV^0_sY^0_1)\Re(\langle V^+\rangle)\Re(1+\zeta^\prime)]\nonumber\\
   &+&|\Pi_{K_2^*}|^2[|\langle T^+\rangle|^2 +(L^+_sT^0_sY^0_2)^2|1+\zeta^\prime|^2
      -2(L^+_sT^0_sY^0_2)\Re(\langle T^+\rangle)\Re(1+\zeta^\prime)]\nonumber\\
   &+&2\Re(\Pi_{K_0^*}\Pi_{K^*}^*)[(L^+_0S^0_0-L^+_sS^0_s)Y^0_0\Re(\langle V^+\rangle )-(L^+_0S^0_0Y^0_0)(L^+_sV^0_sY^0_1)\Re(1+\zeta^\prime)\nonumber\\
          &-&(L^+_sS^0_sY^0_0)\Re(\langle V^+\rangle )\Re(\zeta^\prime)+(L^+_sS^0_sY^0_0)(L^+_sV^0_sY^0_1)|1+\zeta^\prime|^2]\nonumber\\
   &+&2\Im(\Pi_{K_0^*}\Pi_{K^*}^*)\Im(\langle V^+\rangle )[(L^+_0S^0_0-L^+_sS^0_s)Y^0_0-(L^+_sS^0_sY^0_0)\Re(\zeta^\prime)]\nonumber\\
   &+&2\Re(\Pi_{K_0^*}\Pi_{K_2^*}^*)[(L^+_0S^0_0-L^+_sS^0_s)Y^0_0\Re(\langle T^+\rangle )-(L^+_0S^0_0Y^0_0)(L^+_sT^0_sY^0_2)\Re(1+\zeta^\prime)\nonumber\\
          &-&(L^+_sS^0_sY^0_0)\Re(\langle T^+\rangle )\Re(\zeta^\prime)+(L^+_sS^0_sY^0_0)(L^+_sT^0_sY^0_2)|1+\zeta^\prime|^2]\nonumber\\
   &+&2\Im(\Pi_{K_0^*}\Pi_{K_2^*}^*)\Im(\langle T^+\rangle )[(L^+_0S^0_0-L^+_sS^0_s)Y^0_0-(L^+_sS^0_sY^0_0)\Re(\zeta^\prime)]\nonumber\\
   &+&2\Re(\Pi_{K^*}\Pi_{K_2^*}^*)[\Re(\langle V^+\rangle \langle T^+\rangle^*)-(L^+_sT^0_sY^0_2)\Re(\langle V^+\rangle )-(L^+_sT^0_sY^0_2)\Re(\langle V^+\rangle )\Re(\zeta^\prime)\nonumber\\
                &-&(L^+_sV^0_sY^0_1)\Re(\langle T^+\rangle )\Re(1+\zeta^\prime)+(L^+_sV^0_sY^0_1)(L^+_sT^0_sY^0_2)|1+\zeta^\prime|^2]\nonumber\\
   &-&2\Im(\Pi_{K^*}\Pi_{K_2^*}^*)[\Im(\langle V^+\rangle \langle T^+\rangle^*)-(L^+_sT^0_sY^0_2)\Im(\langle V^+\rangle )-(L^+_sT^0_sY^0_2)\Im(\langle V^+\rangle )\Re(\zeta^\prime)\nonumber\\
                &+&(L^+_sV^0_sY^0_1)\Im(\langle T^+\rangle )\Re(1+\zeta^\prime)],
\ee
and the CP-odd quantity ${\cal D}$ is
\be
{\cal D}=-2\Im(\zeta^\prime)C(q^2,s_{_M})\Delta,
\label{calD}
\ee
with
\be
&&\Delta=2\bigg[\Im(\langle V^+\rangle ) 
 \{(L^+_sV^0_sY^0_1)|\Pi_{K^*}|^2+(L^+_sS^0_sY^0_0)\Re(\Pi_{K_0^*}\Pi_{K^*}^*) \nonumber\\
&& ~~~~~~~~+(L^+_sT^0_sY^0_2)\Re(\Pi_{K^*}\Pi_{K_2^*}^*)\}\nonumber\\
&&+\Im(\langle T^+\rangle )\{(L^+_sT^0_sY^0_2)|\Pi_{K_2^*}|^2+(L^+_sS^0_sY^0_0)\Re(\Pi_{K_0^*}\Pi_{K_2^*}^*)+(L^+_sV^0_sY^0_1)\Re(\Pi_{K^*}\Pi_{K_2^*}^*)\}\nonumber\\
&&+\Re(\langle V^+\rangle )\{(L^+_sT^0_sY^0_2)\Im(\Pi_{K^*}\Pi_{K_2^*}^*)-(L^+_sS^0_sY^0_0)\Im(\Pi_{K_0^*}\Pi_{K^*}^*)\}\nonumber\\
&&-\Re(\langle T^+\rangle )\{(L^+_sV^0_sY^0_1)\Im(\Pi_{K^*}\Pi_{K_2^*}^*)+(L^+_sS^0_sY^0_0)\Im(\Pi_{K_0^*}\Pi_{K_2^*}^*)\}\nonumber\\
&&+(L^+_0S^0_0Y^0_0)(L^+_sV^0_sY^0_1)\Im(\Pi_{K_0^*}\Pi_{K^*}^*)+(L^+_0S^0_0Y^0_0)(L^+_sT^0_sY^0_2)\Im(\Pi_{K_0^*}\Pi_{K_2^*}^*)\bigg],
\label{oquantity}
\ee
where
\be
\langle V^\pm\rangle \equiv\sum_{\lambda =0,\pm 1}L^\pm_\l V^\l_\l Y^\l_1,\;\;
\langle T^\pm\rangle \equiv\sum_{\lambda =0,\pm 1}L^\pm_\l T^\l_\l Y^\l_2,
\ee
and the overall function $C(q^2,s_{_M})$ is given by
\be
C(q^2,s_{_M})=|V_{ub}|^2\frac{G_F^2}{2}\frac{1}{2m_D}
       \frac{(q^2-m_l^2)\sqrt{Q_+Q_-}}{256\pi^3m_D^2q^2}.
\ee
Note that the CP-odd term is proportional to the imaginary part 
of the parameter $\zeta$ and the lepton mass.
Therefore, in order to investigate CP violation in the $D_{l4}$ decays, 
we have to consider massive leptonic 
$D_{\mu 4}$ ($D^\pm\to K\pi \mu^\pm \nu$) decays.

\section{Numerical Analyses and Conclusions}

First, we calculate
$D_{\mu 4}$ decay rates through the most dominant resonance, $K^*(892)$,
by varying values of the scalar coupling and then compare those results with
the present experimental branching ratio \cite{pdg},
\be
{\cal BR}[D^+\to(\overline{K^*}(892)^0 \to K^-\pi^+)\mu^+\nu_\mu]
=(2.9\pm 0.4)\%.
\label{dmu4}
\ee
In Fig.~3 we show the allowed parameter space of the complex scalar coupling.
For example, using the present experimental result of Eq. (\ref{dmu4}),
one can constrain the value of the scalar coupling down to
$$|\zeta| \sim 5.0$$ 
at 2--$\sigma$ level.
In our numerical analyses, 
we use the so-called ISGW2 form factors \cite{ISGW2}
in $D\to \tilde{K}_i$ transition amplitudes in Eq.~(\ref{BMamp}). 
\begin{figure}[tb]
\vspace*{-4.2cm}
\hbox to\textwidth{\hss\epsfig{file=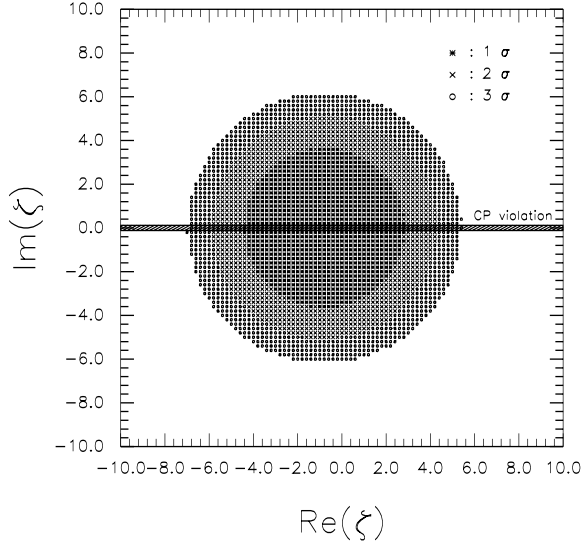,height=16cm,width=12.5cm,angle=0}\hss}
\vspace*{-5.0cm}
\caption{The allowed parameter space of the complex scalar coupling $\zeta$
at each confidence level, by comparing the theoretical decay rate with 
the experimental result, Eq. (\ref{dmu4}), 
for $D^+\to(\overline{K^*}(892)^0 \to K^-\pi^+)\mu^+\nu_\mu$ decay.
The horizontal shaded region represents the attainable 2--$\sigma$ limits on the
imaginary part through the optimal CP-odd asymmetry (see Table 2).}
\label{fig:pspace}
\end{figure}

As  mentioned earlier, study of CP violation effects 
can give a further constraint to 
the imaginary part of the scalar coupling.
We consider the so-called optimal observable.
An appropriate real weight function $w(s_{_M},q^2;\theta^*,\theta_l,\phi^*)$
is usually employed to separate the CP-odd ${\cal D}$ contribution and  enhances
its analysis power through the CP-odd quantity,
\begin{eqnarray}
\langle w{\cal D}\rangle\equiv\int\left[w{\cal D}\right] d\Phi_4 ~.
\end{eqnarray}
And the analysis power is determined by the parameter,
\begin{eqnarray}
\varepsilon
   =\frac{\langle w{\cal D}\rangle}{\sqrt{\langle{\cal S}\rangle
          \langle w^2{\cal S}\rangle}}\;.
\label{Significance}
\end{eqnarray}
For the analysis power $\varepsilon$, the number $N_D$ of the $D$-mesons 
needed to observe CP violation at 1--$\sigma$ level is
\begin{eqnarray}
N_D=\frac{1}{Br\cdot\varepsilon^2}\;.
\label{eq:number}
\end{eqnarray}
From the above relation, we can also deduce the bound on the CP-odd parameter
for given $N_D$ at an arbitrary confidence level.
Certainly, it is desirable to find the optimal weight function
with the largest analysis power. It is known  \cite{Optimal} that 
when the CP-odd contribution to the total rate is relatively small, 
the optimal weight function  is approximately given as
\begin{eqnarray}
w_{\rm opt}(s_{_M},q^2;\theta^*,\theta_l,\phi^*)=
 \frac{{\cal D}}{{\cal S}}~~~\Rightarrow~~~
 \varepsilon_{\rm opt}=\sqrt{\frac{\langle\frac{{\cal D}^2}{{\cal S}}\rangle}
 {\langle{\cal S}\rangle}}.
\end{eqnarray}
We adopt this optimal weight function in our numerical analyses.
\begin{table}
{Table~2}. {Attainable 2--$\sigma$ limits on the imaginary 
             part of the scalar coupling, $\zeta$, through the optimal 
             observable, with the number of $D$-mesons, $N_D=10^8$}\par
\begin{center}
\begin{tabular}{c|c}
\hline
     intermediate states & Attainable 2--$\sigma$ limits \\ \hline
     $K^*$, $K_0^*$ and $K_2^*$ & $|\Im(\zeta)|=0.121$\\
     $K^*$ only & $|\Im(\zeta)|=0.126$\\
     \hline
\end{tabular}
\end{center}
\end{table}

In Table 2, we show the attainable 2--$\sigma$ limits on the imaginary
part of the scalar coupling $\zeta$ through the optimal observable.
In order to estimate effects from higher excited resonances,
we separately present the result obtained by including only the lowest
state $K^*$ as an intermediate state.
We can easily see that in $D_{l4}$ decays the effect
of higher excited resonances is very small.
We note the authors of Ref.~\cite{korner}
analyzed the possibility of probing CP-violation by extracting 
T-odd angular correlations in $D\to K^*(\to K\pi)l\nu$ decay 
and found that the effects can be detected in some cases.

In order to get explicit meaning for our analyses,
we now consider specific models beyond the SM.
As specific extensions of the SM,
we consider four types of scalar-exchange models which preserve the 
symmetries of the SM \cite{scalar};
one of them is the multi-Higgs-doublet (MHD) model \cite{Grossman} and
the other three  are the scalar-leptoquark (SLQ) models \cite{wyler,randall}:
\begin{itemize}
\item
(i) Assuming that all but the lightest of the charged
scalars effectively decouple from fermions, 
the effective Lagrangian of the MHD model contributing 
to the decay $D\to K\pi l\bar{\nu}_l$ 
is  given at energies considerably low compared to $M_H$  by 
\be
{\cal L}_{_{MHD}}=2\sqrt{2}G_FV_{cs}\frac{m_l}{M_H^2}\Big[m_s XZ^*(\bar{c}_L s_R)
            +m_c YZ^*(\bar{c}_R s_L)\Big](\bar{l}_R \nu_L),
\ee
where $X$, $Y$ and $Z$ are complex coupling constants which can be
expressed in terms of the charged Higgs mixing matrix elements.
From the effective Lagrangian for the MHD model, we obtain scalar coupling
$\zeta_{_{MHD}}$   
\be                                            
\zeta_{_{MHD}}=\frac{m_l m_c}{M_H^2}\Big\{(\frac{m_s}{m_c})XZ^*-YZ^*\Big\}.
\label{etaMHD}
\ee
The present bound \cite{Grossman} on $\zeta_{_{MHD}}$ is
\be
|\zeta_{_{MHD}}|\;<\; 0.0029\;\qquad\qquad {\rm for}\;\mu\;{\rm family}.
\label{Hmodel}
\ee
Therefore, the MHD bound is already too stringent to use this $D_{\mu 4}$
decay mode for further constraining the coupling constant.
\item
(ii) The effective Lagrangians for the three SLQ models \cite{scalar,wyler}
contributing to the decay $D\to K\pi l\nu$ are written, 
after a few Fierz rearrangements,  in the form
\be
{\cal L}^{^{I}}_{_{SLQ}}&=&-\frac{x_{2j}x^{\prime *}_{2j}}{2M^2_{\phi_1}}\left[
             (\bar{s}_Lc_R)(\bar{\nu}_{lL}l_R)
            +\frac{1}{4}(\bar{s}_L\sigma^{\mu\nu}c_R)
             (\bar{\nu}_{lL}\sigma_{\mu\nu}l_R)\right]+h.c.,\nonumber\\
{\cal L}^{^{II}}_{_{SLQ}}&=&-\frac{y_{2j}y^{\prime *}_{2j}}{2M^2_{\phi_2}}\left[
             (\bar{s}_Lc_R)(\bar{l}^c_R\nu^c_{l L})
            +\frac{1}{4}(\bar{s}_L\sigma^{\mu\nu}c_R)
          (\bar{l}^c_R\sigma_{\mu\nu}\nu^c_{l L})\right]\nonumber\\
          &&+\frac{y_{2j}y^*_{2j}}{2M^2_{\phi_2}}(\bar{s}_L\gamma_\mu c_L)
           (\bar{l}^c_L\gamma^\mu\nu^c_{l L})+h.c.,\nonumber\\
{\cal L}^{^{III}}_{_{SLQ}}&=&-\frac{z_{2j}z^*_{2j}}{2M^2_{\phi_3}}(\bar{s}_L\gamma_\mu c_L)
           (\bar{l}^c_L\gamma^\mu\nu^c_{l L})+h.c.\;,
\ee
where $j=1,2$ for $l=e,\mu$, respectively, 
and the coupling constants $x^{(\prime)}_{ij}$, $y^{(\prime)}_{ij}$ and
$z_{ij}$ are in general complex numbers so that the CP symmetry is violated
in the scalar-fermion Yukawa interaction terms. 
Then we find that the derived SLQ model
couplings are
\begin{eqnarray}
&&\zeta^{^I}_{_{SLQ}}=-\frac{x_{2j}
      x^{\prime *}_{2j}}{4\sqrt{2}G_FV_{ub}M^2_{\phi_1}}\,,\nonumber\\
&&\zeta^{^{II}}_{_{SLQ}}= -\frac{y_{2j}y^{\prime *}_{2j}}
   {4\sqrt{2}G_FV_{ub}M^2_{\phi_2}}\,,\nonumber\\
&&\zeta^{^{III}}_{_{SLQ}}= 0\,.
\label{LQp}
\end{eqnarray}
Although there are at present no direct 
constraints on the SLQ model CP-odd parameters in (\ref{LQp}), 
rough constraints on the parameters can be obtained by the assumption 
\cite{Davidson} that $|x^\prime_{2j}|\sim |x_{2j}|$ and 
$|y^\prime_{2j}|\sim |y_{2j}|$. That is to say, the leptoquark couplings   
to quarks and leptons belonging to the same generation are of a similar
size. Then the experimental upper bounds 
yield \cite{Davidson}
\begin{eqnarray}
|\zeta^{^I}_{_{SLQ}}|< 6.23,\quad 
|\zeta^{^{II}}_{_{SLQ}}|< 15.56 \qquad\; {\rm for}\;\mu\;{\rm family}.
\label{Models}
\end{eqnarray}
Therefore using these $D_{\mu 4}$ decays, one could extract much more stringent
constraints on $\zeta^{^{I,II}}_{_{SLQ}}$, as shown in Table 2.
\end{itemize}

To summarize, we have investigated new physics effects 
through the semileptonic $D_{\mu 4}$ decays, $D^\pm\to K\pi \mu^\pm \nu_\mu$, 
where we extended the weak charged current 
by including a scalar-exchange interaction with a complex coupling.
We found that by comparing the theoretical decay rate 
with the observed experimental value, 
one can constrain both real and imaginary parts of the complex scalar coupling.
We also investigated the direct CP violation effects,
and found that one can  constrain further the imaginary part
through the CP-odd optimal asymmetry.
We considered as specific models the multi-Higgs doublet model and the scalar
leptoquark models, and found that 
one can extract much more stringent constraints 
on the scalar-leptoquark couplings, $\zeta^{^{I,II}}_{_{SLQ}}$, 
through the decay mode $D^\pm\to K\pi \mu^\pm \nu$.
\\


\section*{Acknowledgments}

\noindent
We thank G. Cvetic for careful reading of the manuscript and 
his valuable comments.
The work of C.S.K. was supported 
in part by  grant No. 1999-2-111-002-5 from the Interdisciplinary 
Research Program of the KOSEF,
in part by the BSRI Program, Ministry of Education, Project No. 98-015-D00061,
in part by KRF Non-Directed-Research-Fund, Project No. 1997-001-D00111, and
in part by the KOSEF-DFG large collaboration project, 
Project No. 96-0702-01-01-2.
J.L. and W.N. wish to acknowledge the financial support of 
1997-sughak program of Korean Research Foundation, Project No. 1997-011-D00015.

%

\end{document}